\documentclass{aa}  

\usepackage{placeins}
\usepackage{natbib}
\usepackage{amsmath}
\usepackage{amsfonts}
\usepackage{amssymb}
\usepackage{graphicx}
\usepackage[titletoc]{appendix}
\usepackage{color}
\usepackage{hyperref}
\usepackage{cleveref}
\usepackage{physics}
\usepackage[rightcaption]{sidecap}
\usepackage{subfigure}
\usepackage{comment}
\usepackage{soul}
\usepackage{cancel}
\usepackage{dcolumn}

\usepackage{array}
\usepackage{ctable}
\usepackage{multirow}
\usepackage{siunitx}
\usepackage{longtable}
\usepackage{tabularx}
\usepackage{booktabs}
\usepackage{float}

\graphicspath{{Graphics/}}

\def\be{\begin{equation}}
\def\ee{\end{equation}}
\def\bea{\begin{eqnarray}}
\def\eea{\end{eqnarray}}

\def\apjl{ApJ Lett.}
\def\apjs{ApJ Suppl. Ser.}

\def\aap{A\&A}

\definecolor{vividviolet}{rgb}{0.62, 0.0, 1.0}
\definecolor{amaranth}{rgb}{0.9, 0.17, 0.31}
\definecolor{palatinateblue}{rgb}{0.15, 0.23, 0.89}
\definecolor{brightpink}{rgb}{1.0, 0.0, 0.5}
\definecolor{cornflowerblue}{rgb}{0.39, 0.58, 0.93}
\definecolor{deepcarminepink}{rgb}{0.94, 0.19, 0.22}
\definecolor{radicalred}{rgb}{1.0, 0.21, 0.37}

\hypersetup{ linktoc=all,
    colorlinks, linkcolor={palatinateblue},
    citecolor={brightpink}, urlcolor={amaranth}
}

\DeclareUnicodeCharacter{2212}{\ensuremath{-}}

\begin{document} 

\title{The macroscopic precession model of quasiperiodic oscillations for rotating compact objects}
\titlerunning{The MPM for rotating compact objects}

\author{Orlando Luongo\inst{1,2,3,6}\corrauth{orlando.luongo@unicam.it}      
\and Marco Muccino\inst{1,4,5,6}\email{marco.muccino@unicam.it}
}
\institute{Universit\`a di Camerino, Via Madonna delle carceri 9, 62032 Camerino, Italy.
\and
Department of Nanoscale Science and Engineering, University at Albany SUNY, Albany, NY 12222, USA.
\and
INAF - Osservatorio Astronomico di Brera, Milano, Italy.
\and
INAF - Catania Astrophysical Observatory, Via S.Sofia 79, 95123 Catania, Italy.
\and
ICRANet, Piazza della Repubblica 10,  65122 Pescara, Italy.
\and 
Al-Farabi Kazakh National University, Al-Farabi av. 71, 050040 Almaty, Kazakhstan.
}

\date{Received Month XX, 20XX}

\abstract 
{The relativistic precession model interprets twin kilohertz quasiperiodic oscillations (QPOs) as geodesic frequencies of test particles orbiting in the space-time of X-ray binaries hosting either a neutron star (NS) or a black hole.}
{In several NS X-ray binaries, QPOs are well reproduced by effective geometries nearly degenerate with a Schwarzschild-de Sitter (SdS) space-time, hindering independent determinations of mass, angular momentum, and other observables. We propose to solve this physical limitation by incorporating orbiting matter spin effects, culminating in the introduction of the macroscopic precession model (MPM).} 
{We treated disk inhomogeneities as spinning test bodies governed by the Mathisson-Papapetrou-Dixon equations and obtained non-minimal spin curvature corrections for the azimuthal and the radial epicyclic frequencies. 
We performed Monte chain Carlo Markov fits, based on the Metropolis algorithm, and modeled eight NS X-ray binary sources.}
{Our statistical analyses show that the data select the internal structure of the accreting matter without requiring corrections to the Kerr and Schwarzschild space-times through the introduction of any de Sitter phase.}
{Physically, the MPM paradigm explains why an effective SdS-like structure could be statistically favored if spin is not employed, through non-minimal spin-curvature coupling, leaving the test particle hypothesis unaltered.}


\maketitle
\nolinenumbers

\section{Introduction}

Twin kilohertz quasiperiodic oscillations (QPOs) are among the most promising strong-gravity probes in accreting compact systems \citep{1998ApJ...492L..59S,vanderklis06}. They track the innermost disk where orbital, radial, and vertical timescales are set by the gravitational field \citep{1999PhRvL..82...17S,belloni14}. 

The well-established relativistic precession model (RPM) paradigm identifies these frequencies with azimuthal and periastron frequencies, describing test particles orbiting compact objects. This provides a promising scenario since it relies solely on standard general relativity, thereby implying de facto geodesic metric explanations for QPOs \citep{1998ApJ...492L..59S}. 
The RPM faces challenges in fitting current data, suggesting that Markov chain Monte Carlo (MCMC) analyses may favor the effective -- though not physically motivated -- Schwarzschild-de Sitter (SdS) metric \citep{boshkayev23b,Boshkayev:2023rhr}. This metric even outperforms the Kerr space-time, providing unphysical neutron star (NS) masses and/or spins \citep{2024MNRAS.531.3876B}. 
The numerical results have been confirmed even at the level of anharmonic orders \citep{Giambo:2025ukm}, suggesting that a phenomenological de-Sitter correction could explain the process, without a clear explanation. 

In this work, we demonstrate that the RPM, although conceptually viable, does not account for the internal structure of test bodies. In this respect, X-ray modulation is rather produced by macroscopic elements in the disk. These include vortices, clumps, or inhomogeneities, which can remain test bodies without being structureless. To this end, we incorporate the Mathisson-Papapetrou-Dixon (MPD) dynamics, where a non-minimal spin-curvature coupling is included \citep{Poisson:2011nh,loomis17,Costa:2017kdr}, and compute the leading order correction while preserving the test particle assumption. 
The corresponding description, dubbed the macroscopic precession model (MPM), does not exactly rely on geodesic motion and furnishes -- in its simplest representation -- the possibility to model QPOs through a further spin contribution. Expanding the spin term up to different orders, we show that, in the case of Kerr geometry, one can forecast highly viable intervals of mass, angular momentum, and frequencies associated with the QPOs for eight NS X-ray binaries. For each of these, we evaluate MCMC analyses, comparing our findings with the Schwarzschild and SdS metrics as well as directly confronting the RPM with our paradigm. We indicate that our scenario predicts viable ranges for the free parameters, which overall outperform those of the RPM. The Kerr case is also used as a tool to correctly predict results that may be biased by relying solely on the RPM. Last but not least, we conclude that, according to our numerical analyses, the RPM does not appear to be viable and may be extended through the spin-curvature interacting term. 

This paper is structured as follows. Section~\ref{sec:2} introduces the theoretical framework of the MPM, which we applied to the Kerr and the Schwarzschild space-times. Section~\ref{sec:3} details our numerical MCMC analyses and summarizes our results. Section~\ref{sec:4} concludes our work, providing perspectives.

\section{Theoretical framework}
\label{sec:2}

Let $p^\mu$ be the momentum, $u^\mu=\dd x^\mu/\dd\tau$ be the kinematic velocity, and $S^{\mu\nu}$ be the antisymmetric spin tensor of a small extended body of mass $m$. The dynamics along the world line, parameterized by the affine parameter $\tau$, is governed by the MPD equations:
\begin{subequations}
\label{eq:mpd}
\begin{align}
 \frac{D p^\mu}{D\tau} &=-\frac12 R^\mu{}_{\nu\alpha\beta}u^\nu S^{\alpha\beta}\ \ ,\ \  p^\mu = m u^\mu + u_\lambda \frac{D S^{\mu\lambda}}{D\tau},\\
 \frac{D S^{\mu\nu}}{D\tau} &=p^\mu u^\nu-p^\nu u^\mu\,.
\end{align}
\end{subequations}
Here, $R^\mu_{\ \nu\alpha\beta}$ is the Riemann tensor, and $D/D\tau$ is the covariant derivative along $\tau$. 
The test body centroid is framed by the Tulczyjew-Dixon (TD) condition, $S^{\mu\nu}p_\nu=0$. Although it is internally structured, this centroid retains the test-particle approximation (TPA) when its gravitational field is negligible:
\begin{equation}
 \kappa=\frac{|S_0|}{m r}\ll1\qquad,
 \qquad S_0^2=\frac12 S_{\mu\nu}S^{\mu\nu}.
 \label{eq:test}
\end{equation}

For any stationary and axisymmetric space-time,
\begin{equation}
ds^2 = g_{tt} dt^2 + g_{rr} dr^2 + g_{\theta\theta} d\theta^2 
+ 2 g_{t\phi} dt d\phi + g_{\phi\phi} d\phi^2\,,
\end{equation}
the Killing charges acquire spin terms,
\begin{equation}
C[\xi]=p^\mu\xi_\mu+\frac12\nabla_\mu\xi_\nu S^{\nu\mu}\,,
\label{eq:killing}
\end{equation}
where $\nabla_\mu$ is the covariant derivative. 
The equatorial circular motion, $v^r=v^\theta=0$, upon the dynamical velocity, $v^\mu=p^\mu/m$, defines the Keplerian frequency, $\Omega_\phi=v^\phi/v^t$.
Thus, the normalization condition, $g_{\mu\nu}v^\mu v^\nu = -1$, gives 
$v^t = (-g_{tt} - 2 g_{t\phi} \Omega_\phi - g_{\phi\phi} \Omega_\phi^2)^{-1/2}$
as well as the effective potential,
\begin{equation}
\mathcal V(r,\theta) = 1 - \frac{
g_{\phi\phi}X^2 + 2 g_{t\phi}XY + g_{tt}Y^2
}{g_{t\phi}^2 - g_{tt} g_{\phi\phi}}\,,
\end{equation}
where, from Eq.~\eqref{eq:killing}, we define
\begin{subequations}
\begin{align}
X = \mathcal E + \Delta \mathcal E &= - v^t \left( g_{tt} + g_{t\phi} \Omega_\phi \right)\,, \\
Y = \mathcal L - \Delta \mathcal L &= v^t \left(g_{t\phi}  + g_{\phi\phi} \Omega_\phi \right)\,.
\end{align}
\end{subequations}
Here, $\mathcal E$ and $\mathcal L$ denote the specific energy and angular momentum, respectively, and $\Delta \mathcal E$ and $\Delta \mathcal L$ their corresponding corrections.

Applying the TD gauge, $\mathcal S^{\mu\nu} v_\nu = 0$, and assuming the extra symmetry condition (ESC), namely the spin of the test body orthogonal to the orbital plane\footnote{In stationary disks, the spin of a test body can be reasonably assumed to align with the rotation axis of the central compact object, i.e., perpendicular to the accretion disk plane.}, we obtain
\begin{subequations}
\label{eq:spin_comp}
\begin{align}
{\rm TD:}\quad \mathcal S^{rt} &= - \mathcal S^{r\phi} \left(\frac{g_{t\phi} + g_{\phi\phi}\Omega_\phi}{g_{tt} + g_{t\phi}\Omega_\phi}\right)\quad,\quad \mathcal S^{t\phi}=0\,, \\
{\rm ESC:}\quad \mathcal S^{t\theta} &= \mathcal S^{r\theta} = \mathcal S^{\theta\phi} = 0\,,
\end{align}
\end{subequations}
where $\mathcal S^{\mu\nu}=S^{\mu\nu}/m$ is the specific spin tensor. 
Based on Eqs.~\eqref{eq:spin_comp}, the specific spin tensor is given by
\begin{equation}
\label{eq:S}
\mathcal S^{\mu\nu}=
\begin{pmatrix}
0 & \mathcal S^{tr} & 0 & 0\\
-\mathcal S^{tr} & 0 & 0 & \mathcal S^{r\phi}\\
0 & 0 & 0 & 0\\
0 & -\mathcal S^{r\phi} & 0 & 0
\end{pmatrix}\,,
\end{equation}
and the spin corrections become
\begin{subequations}
\begin{align}
\label{eq:E0}
\Delta\mathcal E\!&=\!\frac{1}{4}\!\left[\partial_r g_{tt} - \partial_r g_{t\phi}\!\left(\frac{g_{tt} + g_{t\phi}\Omega_\phi}{g_{t\phi} + g_{\phi\phi}\Omega_\phi}\right)\!\right]\!\mathcal S^{tr}\!=\!\Delta \mathcal E_0 \mathcal S^{tr},\\
\label{eq:L0}
\Delta\mathcal L\!&=\!\frac{1}{4}\!\left[\partial_r g_{t\phi} - \partial_r g_{\phi\phi}\!\left(\frac{g_{tt} + g_{t\phi}\Omega_\phi}{g_{t\phi} + g_{\phi\phi}\Omega_\phi}\right)\!\right]\!\mathcal S^{tr}\!=\!\Delta \mathcal L_0 \mathcal S^{tr},
\end{align}
\end{subequations}
where $\Delta \mathcal E_0$ and $\Delta \mathcal L_0$ encompass a factor of $1/4$ and the brackets from Eqs.~\eqref{eq:E0} and \eqref{eq:L0}, respectively.
In the spinless limit, $\mathcal S^{\mu\nu}\rightarrow0$, one recovers $v^\mu\rightarrow u^\mu$, $\Delta \mathcal E\rightarrow0$, and $\Delta \mathcal L\rightarrow0$.

At this stage, for stable circular equatorial orbits with $x_0=(r_0,\theta_0)=(r_0,\pi/2)$, and given that $\mathcal E$ and $\mathcal L$ are constants of motion, the modified Keplerian angular frequency can be obtained by solving $\partial_r\mathcal V|_{x_0}=0$. This can be further simplified using the TPA condition in Eqs.~\eqref{eq:test} and~\eqref{eq:spin_comp}, i.e.,
\begin{equation}
\label{eq:TPA2}
\kappa = \frac{|\mathcal S^{tr}|\sqrt{g_{rr}(g_{tt}g_{\phi\phi}\!-\!g_{t\phi}^2) (g_{tt}\!+\!2g_{t\phi}\Omega_\phi\!+\!g_{\phi\phi}\Omega_\phi^2)}}{r|g_{t\phi} + g_{\phi\phi}\Omega_\phi|},
\end{equation}
which leads to 
\begin{align}
\nonumber
&\left[2 s \kappa r
\frac{\partial_r \Delta \mathcal E_0 + \Omega_\phi \partial_r \Delta \mathcal L_0 + (\Delta \mathcal E_0 + \Omega_\phi \Delta \mathcal L_0) \partial_r \ln|\mathcal S^{tr}|}{|g_{t\phi} + g_{\phi\phi}\Omega_\phi|^{-1}\sqrt{g_{rr}(g_{t\phi}^2 -g_{tt}g_{\phi\phi})}}\right]_{x_0}\\
\label{eq:dVr2}
&+\left[\partial_r g_{tt} + 2 \Omega_\phi \partial_r g_{t\phi} + \Omega_\phi^2 \partial_r g_{\phi\phi} \right]_{x_0}=0\,,
\end{align}
where $s$ is the sign of $\mathcal S^{tr}$.
It is evident that the second line of Eq.~\eqref{eq:dVr2} gives the well-known Keplerian angular frequency for axisymmetric space-times, while the first line gives the small perturbation $\propto \kappa$ that preserves the TPA validity.

Subsequently, we expand $\mathcal V(r,\theta)$ up to the second order around $x_0$ and use the stability condition, $\mathcal V(x_0)=\partial_r\mathcal V|_{x_0} = \partial_\theta \mathcal V|_{x_0}=0$. Focusing on small radial ($r$) displacements, we obtain a harmonic equation from which we define the radial angular frequency:
\begin{equation}
\label{eq:omegas}
\Omega^2_{r} = - \frac{\partial_r^2 \mathcal V|_{x_0}}{2 g_{rr} (v^t)^2}\,.
\end{equation}

\subsection{The case of the Kerr space-time}

We specialize all the above to the Kerr space-time:
\begin{subequations}
\label{gKerr}
\begin{align}
&g_{tt} =  - \left(1 - \frac{2 M r}{\Sigma}\right)\ ,\ g_{rr} = \frac{\Sigma}{\Delta}\ ,\ g_{\theta\theta} = \Sigma,\\ 
&g_{t\phi} = - \frac{2 a M r}{\Sigma} \sin^2\theta\ ,\ g_{\phi\phi} = \left(r^2 + a^2 
+ a g_{t\phi} \right) \sin^2\theta,
\end{align}
\end{subequations}
where $a = j M$ is the specific angular momentum, $M$ is the compact object mass, $j$ is the spin parameter, and $\Sigma = r^2 + a^2 \cos^2\theta$ and 
$\Delta = r^2 - 2 M r + a^2$ are auxiliary functions.

In the innermost regions of the accretion disk, the TPA condition in Eq.~\eqref{eq:TPA2} reduces to
\begin{equation}
\label{eq:TPA3}
\kappa_K = \frac{|\mathcal S^{tr}|}{r}\frac{\sqrt{(r-3M \pm 2 a r \omega_0)/M}}{|1+(a/r)^2+\mp 2 a \omega_0|}\ll1\,,
\end{equation}
where $\omega_0=\sqrt{M/r^3}$ is the RPM Schwarzschild angular Keplerian frequency.
Using Eq.~\eqref{eq:TPA3}, we solve Eq.~\eqref{eq:dVr2} perturbatively at the first order, finding
\begin{equation}
\label{eq:sol_per}
\Omega_{\phi,K} = \omega_\phi + \kappa_K \Delta\omega_\phi + \mathcal O(\kappa_K^2)\,,\\
\end{equation}
where the lowest order contribution $\omega_\phi$ is given by
\begin{equation}
\omega_\phi = \frac{\pm\omega_0}{1\pm a \omega_0}\,,
\end{equation}
for the co- ($+$) and counterrotating ($-$) orbits. The full expression of the first-order term $\Delta\omega_\phi$ is given in Appendix~\ref{app:A}.

Finally, from Eq.~\eqref{eq:omegas} we obtain 
\begin{equation}
\label{eq:sol_Or}
\Omega_{r,K}^2 = \omega_r^2 + \kappa_K\Delta\omega_r^2 + \mathcal{O}(\kappa_K^2)\,,
\end{equation}
where at the lowest order we have
\begin{equation}
\omega_r^2 = \omega_\phi^2 \left(1-\frac{6 M}{r} \pm 8 a \sqrt{\frac{M}{r^3}}-\frac{3 a^2}{r^2}\right)\,.
\end{equation}
The first-order term $\Delta\omega_r^2$ is described in Appendix~\ref{app:A}.

With the above definitions, the MPM -- like the RPM -- identifies the lower and upper QPO frequencies with the periastron and Keplerian frequencies, respectively:
\begin{equation}
\label{QPO_freqs}
f_{L,K}=\frac{1}{2\pi}(\Omega_{\phi,K}-\Omega_{r,K})\qquad,\qquad
f_{U,K}=\frac{1}{2\pi}\Omega_{\phi,K}\,.
\end{equation}

We now specify the specific spin tensor by embedding the symmetry of the accretion flow through the ansatz:
\begin{equation}
\label{eq12}
\mathcal S^{tr} = \mathcal C_n r^n\,.
\end{equation}
Qualitatively, this scaling may be associated with filament-like ($n=1$), thin-disk ($n=2$), and thick-disk or quasi-spherical ($n=3$) configurations, respectively.
The amplitude, $\mathcal C_n\ll1$, guarantees that the TPA shall hold across the disk. Observationally, this allows us to identify an upper (lower) limit, $r_{\rm in}$ ($r_{\rm out}$), on the inner (outer) disk edge. 
Using the measured source upper frequencies, $f_{U,k}\pm\sigma_{U,k}$, we identify
\begin{subequations}
\begin{align}
&{\rm max}(f_{{\rm U},k})\equiv f_{U,K}(M,a,\mathcal C_n,n,r_{\rm in}),\\  
&{\rm min}(f_{{\rm U},k}) \equiv f_{U,K}(M,a,\mathcal C_n,n,r_{\rm out})\,,
\end{align}
\end{subequations}
which leads to radii satisfying the condition $r_{\rm ISCO}\leq r_{\rm in}\leq r_{\rm out}$. Here, $r_{\rm ISCO}$ is the radius of the innermost stable circular orbit (ISCO), determined by solving $\Omega_r^2(r_{\rm ISCO})=0$. The true boundary of the disk is observationally precluded. However, Eq.~\eqref{eq:TPA3} allows us to derive an absolute upper limit for it by assuming that the disk properties remain unaltered up to the radius $r_{\rm disk}$, where $\kappa_K\simeq 0.2$. This introduces the radial ordering condition (ROC):
\begin{equation}
\label{eqROC2}
r_{\rm ISCO}\leq r_{\rm in}\leq r_{\rm out}\leq r_{\rm disk}\,,
\end{equation}
where the outer scale is derived from Eq.\eqref{eq:TPA3} at a sufficiently large radius where the TPA ceases to hold, i.e., at $r_{\rm disk}\simeq |\mathcal C_n|^{-1/2}$.

\subsection{The case of the Schwarzschild space-time}

It easy to show that all the formulas reduce to the Schwarzschild case considered in \citet{Bianchini:2025zvp}, when $g_{t\phi}\rightarrow0$ (or equivalently, $a\rightarrow0$).
In this limit, Eqs.~\eqref{gKerr} provide
\begin{equation}
\label{gS}
g_{tt} = - g_{rr}^{-1} = - \left(1 - \frac{2 M}{r}\right)\quad,\quad g_{\theta\theta} = \frac{g_{\phi\phi}}{\sin^2\theta} = r^2\,,
\end{equation}
and the TPA condition in Eq.~\eqref{eq:TPA3} reduces to
\begin{equation}
\label{eq:TPA4}
\kappa_S = \frac{|\mathcal S^{tr}|}{r}\sqrt{\frac{r-3M}{M}}\ll1\,.
\end{equation}
In this case, the TPA holds across the disk when $\kappa_S\lesssim 0.2$. This yields the ROC, as shown with Eq.~\eqref{eqROC2}.

Again, solving Eq.~\eqref{eq:dVr2} perturbatively leads to
\begin{equation}
\label{eq:sol_perS}
\Omega_{\phi,S} = \pm \omega_0 + \kappa_S \Delta\omega_0 + \mathcal O(\kappa_S^2)\,,\\
\end{equation}
for co- ($+$) and counterrotating ($-$) orbits. The value $\Delta\omega_0$ is described in Appendix~\ref{app:A}.
Subsequently, from Eq.~\eqref{eq:omegas} we obtain 
\begin{equation}
\label{eq:sol_OrS}
\Omega_{r,S}^2 = \varpi_r^2 + \kappa_S\Delta\varpi_r^2 + \mathcal{O}(\kappa_S^2)\,,
\end{equation}
where at the lowest order we have
\begin{equation}
\varpi_r^2 = \omega_0^2 \left(1-\frac{6 M}{r} \right)\,.
\end{equation}
The first-order term $\Delta\varpi_r^2$ is described in Appendix.~\ref{app:A}.

Here, we adopt the parameterization from Eq.~\eqref{eq12} and apply the ROC to the observational data through the definition of the lower and the upper QPO frequencies, respectively:
\begin{equation}
\label{QPO_freqsS}
f_{L,S}=\frac{1}{2\pi}(\Omega_{\phi,S}-\Omega_{r,S})\qquad,\qquad
f_{U,S}=\frac{1}{2\pi}\Omega_{\phi,S}\,.
\end{equation}

\section{Numerical analysis and results}
\label{sec:3}

For each source, we considered a set of $N$ twin kilohertz frequency pairs with errors, i.e., $(f_{L,k}\pm\sigma_{L,k},f_{U,k}\pm\sigma_{U,k})$, and identified the best-fit parameters as the set that maximizes the $\ln \bar L$ values of the log-likelihood (LLH) function:
\begin{equation}
\label{eqloglike}
\ln L\!=\!-\frac{1}{2}\!\sum_{k=1}^{N}
\left\{\!
\frac{[f_{L,k}-f_{L,i}(h_i,f_{U,k})]^2}{\sigma_{L,k}^2}
+ \ln\left(2\pi\sigma_{L,k}^2\right)\!
\right\}\!,
\end{equation}
where $i=\{K,S\}$ depends on whether Kerr or Schwarzschild space-times are considered. 
The dependency of $f_{L,i}$ on the radii $r_k$ was shifted to the observed upper frequencies $f_{U,k}$ using the identity\footnote{In general, not all the models have analytic expressions for $f_{L,k} =f_{L,i}(h_i,f_{U,k})$. Thus, $r_k$ is typically evaluated numerically.} $f_{U,k}=f_{U,i}(h_i,r_k)$. 
The model parameters in Eq.~\eqref{eqloglike} are indicated by $h_i=\{M,j_i,\mathcal C_n, n\}$, where the values of the spin parameter are $j_i=\{j,0\}$ for the Kerr and Schwarzschild space-times, respectively.
The parameter space was sampled by means of a Metropolis-Hastings algorithm-based MCMC simulation with $\mathcal O(10^5)$ iterations. 
We adopted uniform broad priors on $M$ and $\mathcal C_n$ and, for the Kerr space-time, Gaussian-distributed priors $\mathcal N(\mu_j,\sigma_j)$ on $j$ with a mean ($\mu_j$) and standard deviation ($\sigma_j$):   
\begin{subequations}
\begin{align}
M&\in[0,5]\,M_\odot,\\
\mathcal C_n&\in[-0.1,0.1]\,{\rm km}^{1-n},\\ 
j&\in \mathcal N(0,1/3)\,.
\end{align}
\end{subequations}
Here, $\sigma_j=1/3$ ensures that $j\in[-1,1]$ at roughly the $3$$\sigma_j$ level. We fixed the power-law index to $n=\{1,2,3\}$ and assessed the best choice on a case-by-case basis.

The best-fit model among each proposed scenario must satisfy both statistical and physical selection criteria.
Statistically, the best model is established by computing the Bayesian evidence, $\ln B_i$, based on the modified Jeffreys’ scale \citep{2008ConPh..49...71T}. Based on the differences $\Delta = \ln B_i-\ln B_0$, where $\ln B_0 = {\rm min}\{\ln B_i\}$, a model is interpreted as weakly excluded if $0\leq\Delta\leq 1$, mildly excluded if $1<\Delta\leq 3$, strongly excluded if $3\leq\Delta\leq5$, or decisively excluded if $\Delta>5$.

In addition to the statistical selection, each model must conform to physical requirements. Neutron stars shall not exceed theoretical upper limits on:
\begin{itemize}
\item[-] the mass, i.e.,  $M\leq3.2$~M$_\odot$ \citep{2002BASI...30..523S}, and  
\item[-] the spin parameter $|j|\lesssim 0.7$ \citep{2011ApJ...728...12L}.
\end{itemize}
Regarding the disk physics, the inferred radii must be constrained to satisfy the ROC, namely $r_{\rm ISCO}\leq r_{\rm in}\leq r_{\rm out}\leq r_{\rm disk}$.

The final outcomes are shown in Appendix~\ref{app:B}.
Figure~\ref{fig:freq} displays the fits of the lower-upper frequency pairs $(f_L,f_U)$ of the eight NS X-ray binaries considered in this work. 
Table~\ref{tab:results} sums up the overall results. 
Results for RPM-S, RPM-SdS \citep{boshkayev23b}, RPM-K \citep{2024MNRAS.531.3876B}, and MPM-S \citep{Bianchini:2025zvp} were obtained with the same pipeline utilized for MPM-K. 
For RPM-SdS we adopted an effective $n=3$ to ensure the correct physical dimensions of the cosmological constant, $R_0$ \citep{boshkayev23b}, as discussed in \citet{Bianchini:2025zvp}.
Several robust features emerge:
\begin{itemize}
\item[-] RPM-S. This model is strongly disfavored ($\Delta\gg 6$), in line with the conclusions of \citet{boshkayev23b} and \citet{Bianchini:2025zvp}.
\item[-] RPM-K. This model is statistically preferred but discarded for Sco~X1, 4U1608--52, and 4U0614+091, as well as for 4U1728–34 (where it is only weakly viable), because the inferred values of $M$ and $j$ are unphysical for NSs. Only in the case of GX~340+0 is the model both statistically and physically weakly viable. 
\item[-] RPM-SdS and MPM-S. The MPM-S model outperforms the RPM-SdS fits ($\Delta>6$) in five cases (Cir~X1, GX~17+2, Sco~X1, 4U1608--52, and 4U0614+091) and is statistically equivalent to it ($\Delta\leq 3$) in the remaining three \citep{Bianchini:2025zvp}. Recalling the cases where the RPM-K is physically discarded, the MPM-S model provides the best fit in seven sources, with the exception of Cir~X1, where the best fit is given by the MPM-K model.
The preference for MPM-S over RPM-SdS reflects that the former describes accretion at the microphysical level, unlike the latter, which is an effective model. In this respect, MPM-S exhibits remarkable clustering in six out of eight sources around disk-like (or thin-disk) configurations with $n=2$, while the remaining two (GX~5-1 and GX~17+2) favor sphere-like (or thick-disk) configurations with $n=3$.
Interestingly, accounting for spinning test particles reduces the index $n=3$ of the effective RPM-SdS model to the more commonly preferred $n=2$ of the MPM-S case. 
\item[-] MPM-K. This model, which includes the rotation of the central compact object, never improves the fits, with the sole exception of the source Cir~X1. However, for this source, the angular momentum is consistent with $j=0$, and $r_{\rm disk}$ is unconstrained, indicating a degeneracy with the simplest MPM-S model.
Remarkably, all the fits resolve the unphysical values of $M$ and $j$ that arise when the RPM model is considered.
Although the Kerr space-time is not preferred over the Schwarzschild one within the MPM framework, it is clear that including the relativistic spin of test particles in the model makes it more complete and self-consistent. 
The signature of this improvement is a clustering toward a preferred value of $n=1$ (in five out of eight cases), which is shifted from the value $n=2$ of the MPM-S due to the addition of the NS angular momentum to the parameter set.
\end{itemize}

\section{Conclusion}
\label{sec:4}

We proposed a novel model that revises the interpretation of kilohertz QPOs, typically addressed using the widely adopted RPM. To do so, we demonstrated that spin-curvature coupling is necessary to ensure that frequencies trace the motion of extended luminous structures in the underlying space-time. 

Specifically, the compact object's metric provides the background, while the disk supplies the macroscopic spin charge that enters the MPD dynamics. Hence, the MPM ensures consistency with general relativity while retaining the test body approximation. Most importantly, it replaces an effective cosmological term with the aforementioned spin-curvature interaction.

There are two immediate falsifiability tests associated with our novel paradigm. First, a source requiring $\kappa\gtrsim1$ in the emitting region would rule out the present interpretation, because the signal could not be produced by test bodies. Second, a robust Kerr preference with physical NS masses and well-constrained $j$ values would indicate measurements of central rotation. The current sample suggests a universal macroscopic precession law, highlighting the overall validity of incorporating test particle spin.

Remarkably, we ruled out the use of RPM-S, since it is consistently statistically disfavored, and RPM-K, since it is statistically and/or physically unsuitable. Furthermore, as shown in Fig.~\ref{fig:freq}, we demonstrated that the best-suited results -- whether accounting for test particle spin or not -- are provided by MPM-S, RPM-SdS, and the newly introduced MPM-K, which incorporates the rotation of the central object.

All our curves, which reproduce each lower-upper frequency pair $(f_L,f_U)$, are nearly coincident with the data distributions. This implies that only the statistical analyses presented above can resolve the puzzle, mostly in favor of the MPM-S framework. Physically, we interpret this finding as evidence that the NS rotations appear small and quite compatible with the static, spherically symmetric case, although we do not exclude the role of the Kerr space-time itself.

Given the robustness of our MPM paradigm, our future work will leverage it to predict novel outcomes related to black holes. This approach offers a new strategy for better constraining free parameters in strong-gravity regimes and, accordingly, for probing the limits of general relativity in domains where Einstein's gravity may break down.

\begin{acknowledgements}
The authors acknowledge Gabriele Bianchini, Roberto Giambò, Sara Elisa Motta and Hernando Quevedo for discussions on the topic of QPOs and Kuantay Boshkayev for debates on compact object physics. OL thanks the hospitality of the National Centre for Nuclear Research (NCBJ), Warsaw, where this work was carried out.
\end{acknowledgements}

\bibliographystyle{aa}
\bibliography{bibliography}

\begin{appendix}

\onecolumn

\section{Spin-induced corrections to the epicyclic frequencies}
\label{app:A}

For the Kerr space-time, the full expression of the first-order correction of the Keplerian frequency in Eq.~\eqref{eq:sol_per} reads
\begingroup
\tiny
\begin{equation}
\label{App:1}
\Delta\omega_\phi\!=\!\mp\frac{sr^2}{8}\!\left\{\!\frac{2\partial_r \mathcal S^{rt}}{\mathcal S^{rt}}\!\left[\frac{r^2 (r-3 M) + a^2 (r-7 M) \pm 2 a r \omega_0 (a^2+3 M r)}{r^5 + a^4 M + a^2 r^2 (r-3 M) \pm 2a^3 r^2 \omega_0(r-M)}\right]\!+\!\frac{r^2 (6 M+r)\!+\!a^2 (5 r-4 M) \pm a r \omega_0 [a^2+3 r (2 M-5 r)]}{r[r^5 + a^4 M + a^2 r^2 (r-3 M) \pm 2a^3 r^2\omega_0 (r-M)]}\!\right\}\!,
\end{equation}
\endgroup
whereas the full expression of the first-order correction of the squared radial epicyclic frequency in Eq.~\eqref{eq:sol_Or} is given by
\begingroup
\tiny
\begin{align}
\nonumber
\Delta\omega_r^2 =& \frac{4\Delta\omega_\phi}{r^2}\left\{a \omega_0^2 (3r^2+a^2-a^3\omega_\phi) - \frac{\omega_\phi}{2r} \left[a^2(r+10 M) - 3 r^2 (r-2 M) \right] \right\}\\
\nonumber
&\mp s a \omega_\phi^2 \left\{\frac{3 a^6+a^4 r (212 M-113 r)+a^2 r^2 (556 M^2-288 M r+101 r^2)+r^4 (228 M^2-340 M r+153 r^2)}{8r^3 (r^2+a^2 \mp 2 a r^2 \omega_0)^2}\right.\\
\nonumber
&\left.+\left(\frac{\partial_r \mathcal S^{rt}}{\mathcal S^{rt}} \right) \frac{a^6+a^4 r (20 M-31 r)+a^2 r^2 (52 M^2-24 M r-41 r^2)+r^4 \left(60 M^2-28 M r-9 r^2\right)}{2r^2(r^2+a^2 \mp 2 a r^2 \omega_0)^2} \right.\\
\nonumber
&\left.+ \left(\frac{\partial_r^2 \mathcal S^{rt}}{\mathcal S^{rt}}\right)\frac{a^6+a^4 r (-2 M+r+10)-a^2 r^2 [M (20-6 r)+(r-10) r]-r^4 (12 M^2-8 M r+r^2)}{r(r^2+a^2 \mp 2 a r^2 \omega_0)^2}\right\}\\
\nonumber
&+\frac{s\omega_\phi^2}{\omega_0}\!\left\{\! \frac{3 a^6 (14 M-5 r)+a^4 r (488 M^2-288 M r+43 r^2)+a^2 r^2 (264 M^3+44 M^2 r-110 M r^2+67 r^3)+r^5 (28 M r+9 r^2\!-\!60 M^2)}{8 r^4 (r^2 + a^2 \mp 2 a r^2 \omega_0)^2}\right.\\
\nonumber
&\left. +\left(\frac{\partial_r \mathcal S^{rt}}{\mathcal S^{rt}} \right) \left[\frac{a^6 (6 M-5 r)+a^4 r \left(44 M^2-64 M r-7 r^2\right)+a^2 r^2 (24 M^3+68 M^2 r-74 M r^2+r^3)+r^5 (24 M^2-20 M r+3 r^2)}{2 r^3 (r^2 + a^2 \mp 2 a r^2 \omega_0)^2}\right]
\right.\\
\label{App:2}
& \left.- \left(\frac{\partial_r^2 \mathcal S^{rt}}{\mathcal S^{rt}}\right) \left[\frac{a^6 (r-11 M)+a^4 r (10 M^2-23 M r+3 r^2)+a^2 r^2 (24 M^3+8 M^2 r-17 M r^2+3 r^3)+r^5 (6 M^2-5 M r+r^2)}{2 r^2 (r^2 + a^2 \mp 2 a r^2 \omega_0)^2}\right]\right\}.
\end{align}
\endgroup
In both Eqs.~\eqref{App:1}-\eqref{App:2} the upper (lower) sign corresponds to  co-rotating (counter-rotating) orbits.

Conversely, for the Schwarzschild space-time the first-order correction of the Keplerian frequency in Eq.~\eqref{eq:sol_perS} simplifies as
\begin{equation}
\label{App:1S}
\Delta\omega_0 = \mp\frac{s}{8r} \left[1+ \frac{6M}{r} + \frac{2\partial_r \mathcal S^{rt}}{\mathcal S^{rt}} (r-3 M)\right]\,,
\end{equation}
and the first-order correction of the squared radial epicyclic frequency in Eq.~\eqref{eq:sol_OrS} is reduced to
\begingroup
\tiny
\begin{equation}
\label{App:2S}
\Delta\varpi_r^2 = \omega_0\left\{\frac{6\Delta\omega_0}{r} (r-2 M)
+ s \left[ \frac{(28 M r+9 r^2 - 60 M^2)}{8 r^3 } +\left(\frac{\partial_r \mathcal S^{rt}}{\mathcal S^{rt}} \right) \frac{(24 M^2-20 M r+3 r^2)}{2 r^2} - \left(\frac{\partial_r^2 \mathcal S^{rt}}{\mathcal S^{rt}}\right) \frac{(6 M^2-5 M r+r^2)}{2 r}\right]\right\}.
\end{equation}
\endgroup
Again, in both Eqs.~\eqref{App:1S}-\eqref{App:2S} the upper (lower) sign corresponds to  co-rotating (counter-rotating) orbits.

In the above Eqs.~\eqref{App:1}-\eqref{App:2S}, the terms $\partial_r \mathcal S^{rt}/\mathcal S^{rt}$ and $\partial_r^2 \mathcal S^{rt}/\mathcal S^{rt}$ simplify when using the essential parameterization of Eq.~\eqref{eq12}, leading to 
\begin{subequations}
\label{Sn}
\begin{align}
\left(\frac{\partial_r \mathcal S^{rt}}{\mathcal S^{rt}} \right)&=\frac{n}{r}\,,\\
\left(\frac{\partial_r^2 \mathcal S^{rt}}{\mathcal S^{rt}} \right)&=\frac{n(n-1)}{r^2}\,.
\end{align}
\end{subequations}
Substituting the two positions from Eq.~\eqref{Sn} into  Eqs.~\eqref{App:1S}-\eqref{App:2S}, we recover the results shown in \citet{Bianchini:2025zvp}.

\section{Results of the MCMC analyses on the NS X-ray binaries}
\label{app:B}

\begin{figure}[t]
{
\includegraphics[width=0.49\hsize,clip]{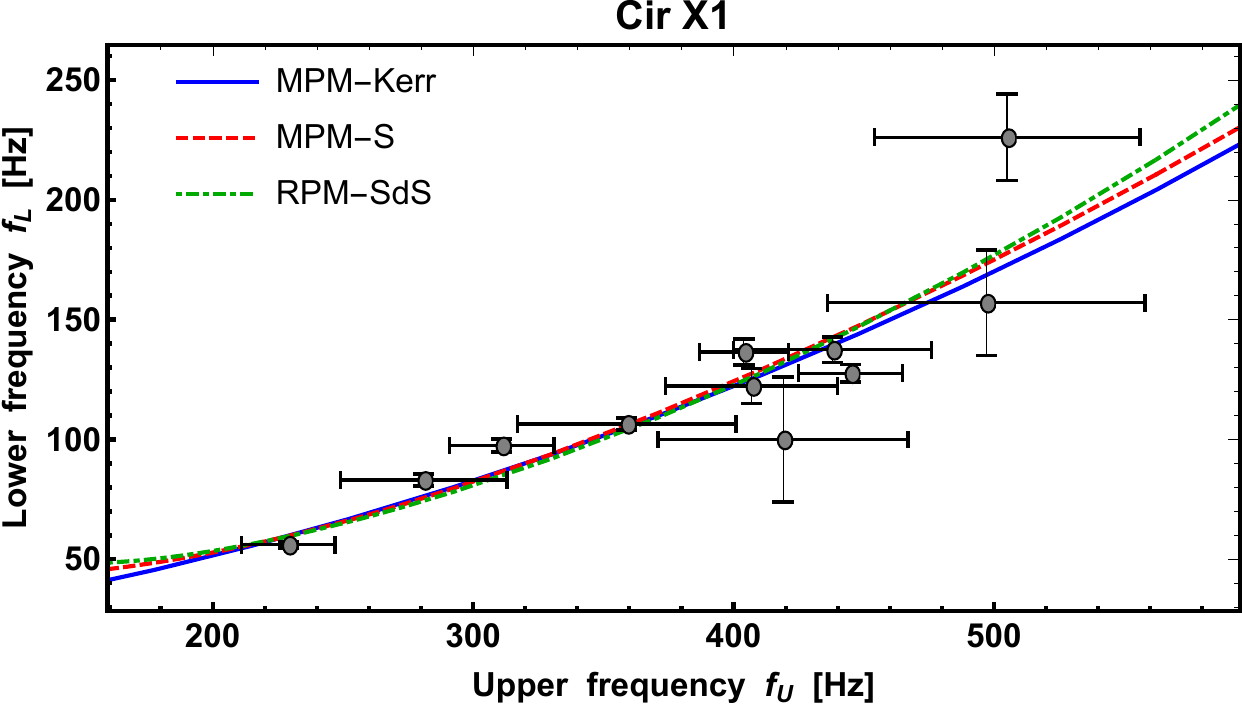}\hfill
\includegraphics[width=0.49\hsize,clip]{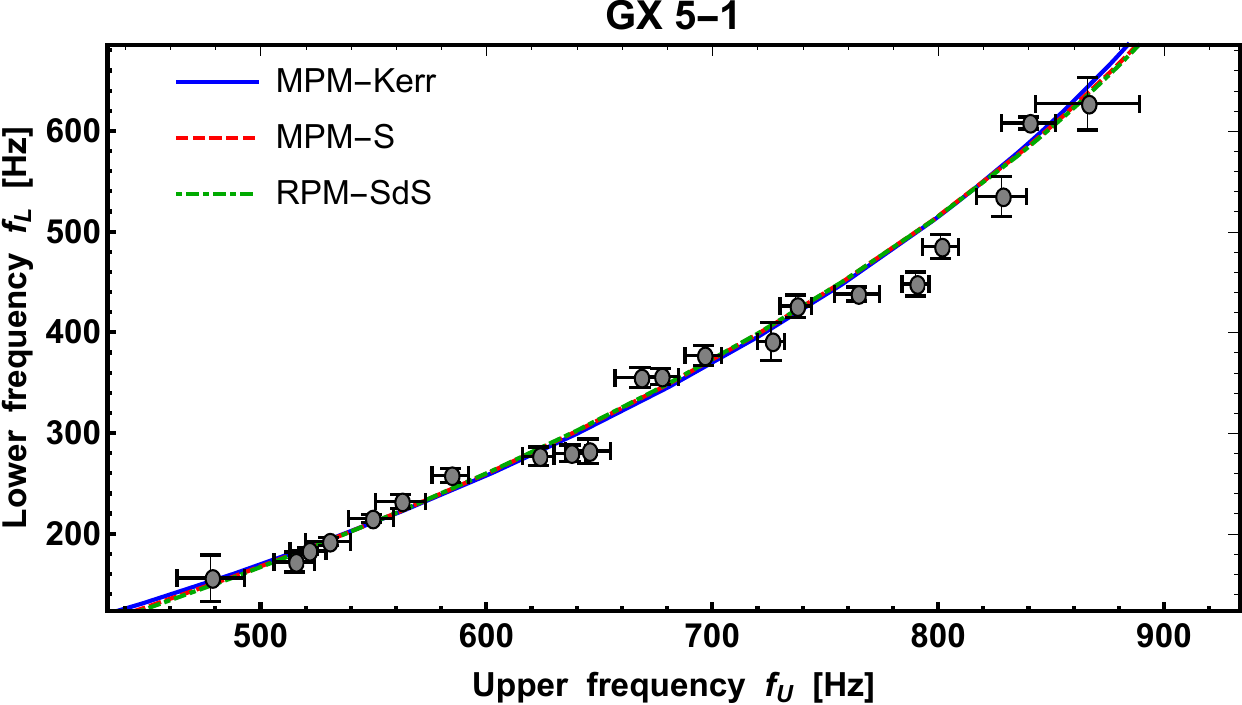}
}
\vspace{0.15cm}

{
\includegraphics[width=0.49\hsize,clip]{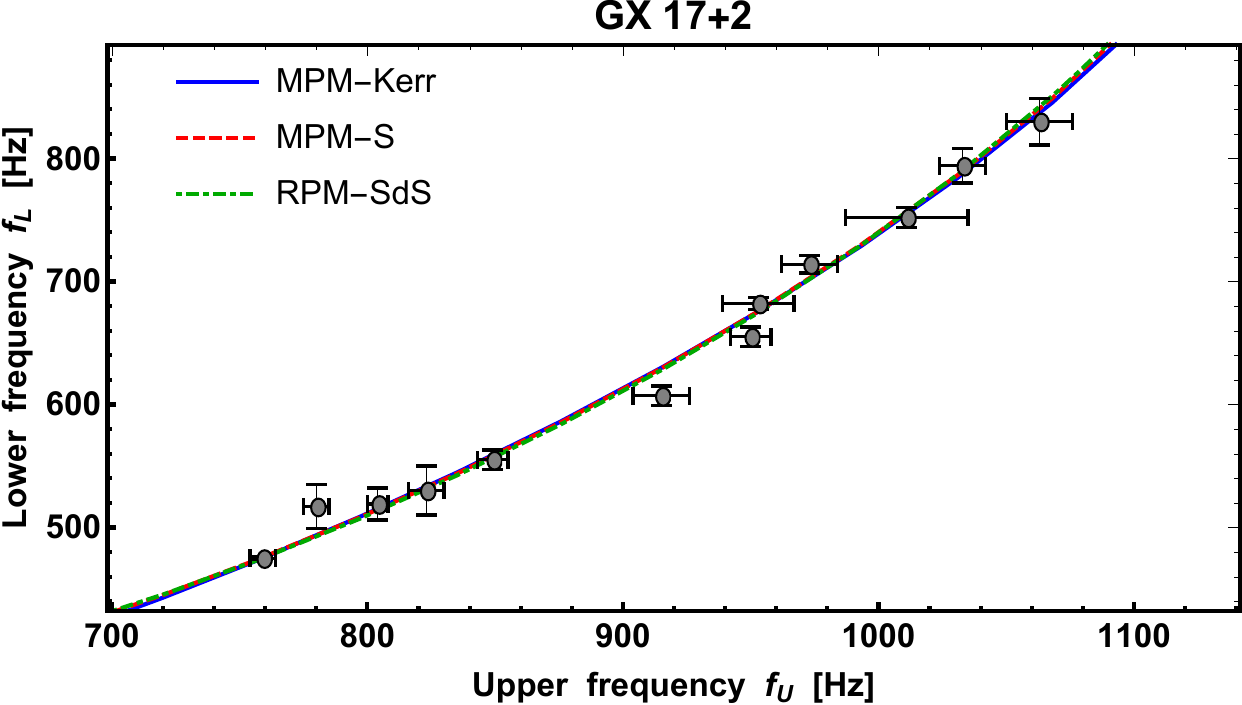}\hfill
\includegraphics[width=0.49\hsize,clip]{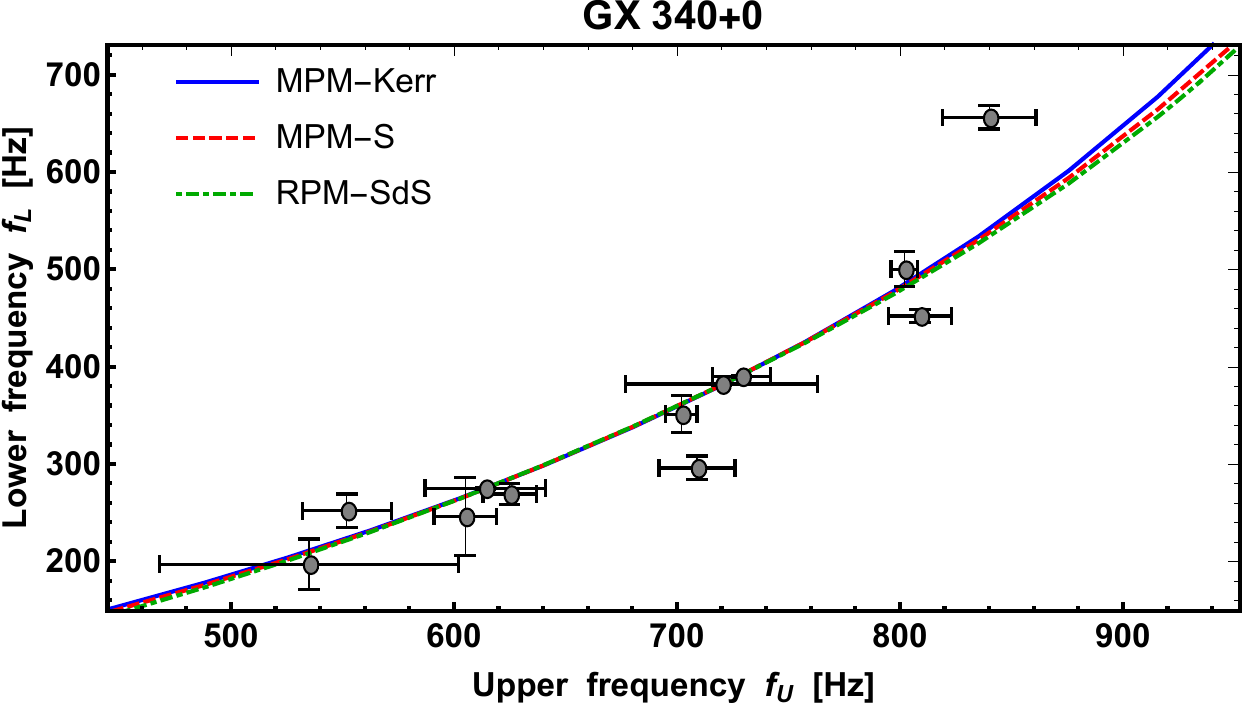}
}

\vspace{0.15cm}
{
\includegraphics[width=0.49\hsize,clip]{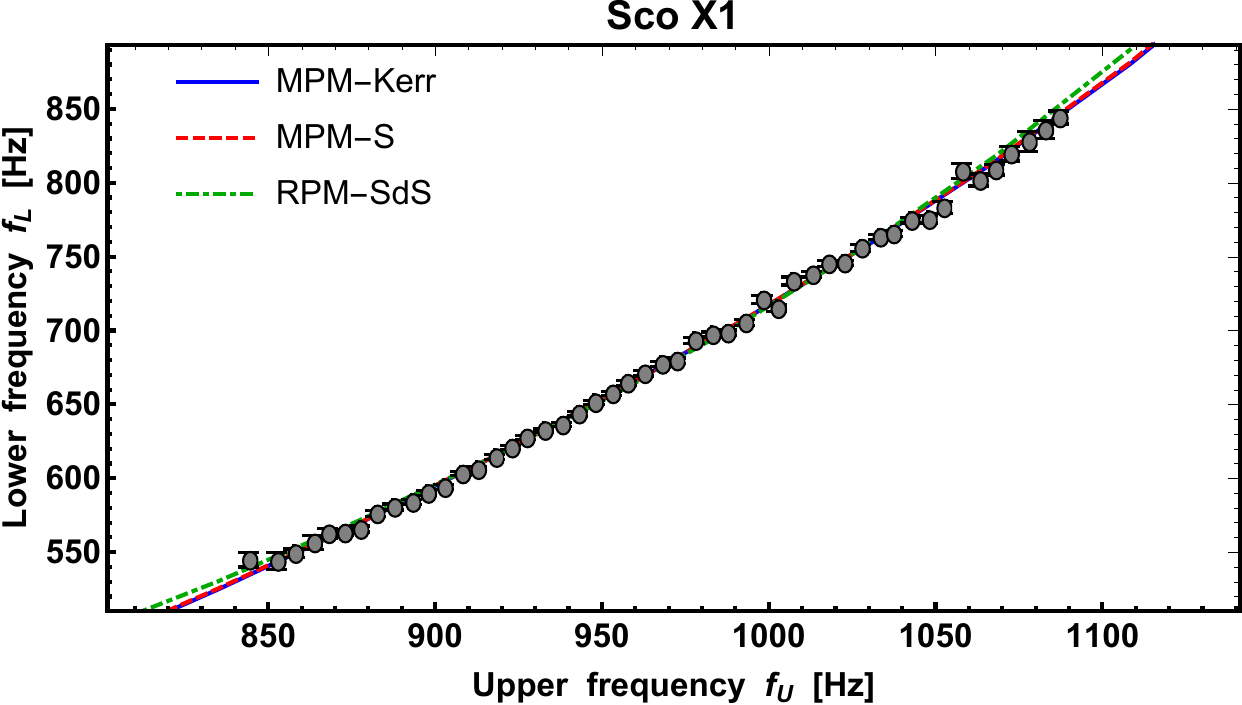}\hfill
\includegraphics[width=0.49\hsize,clip]{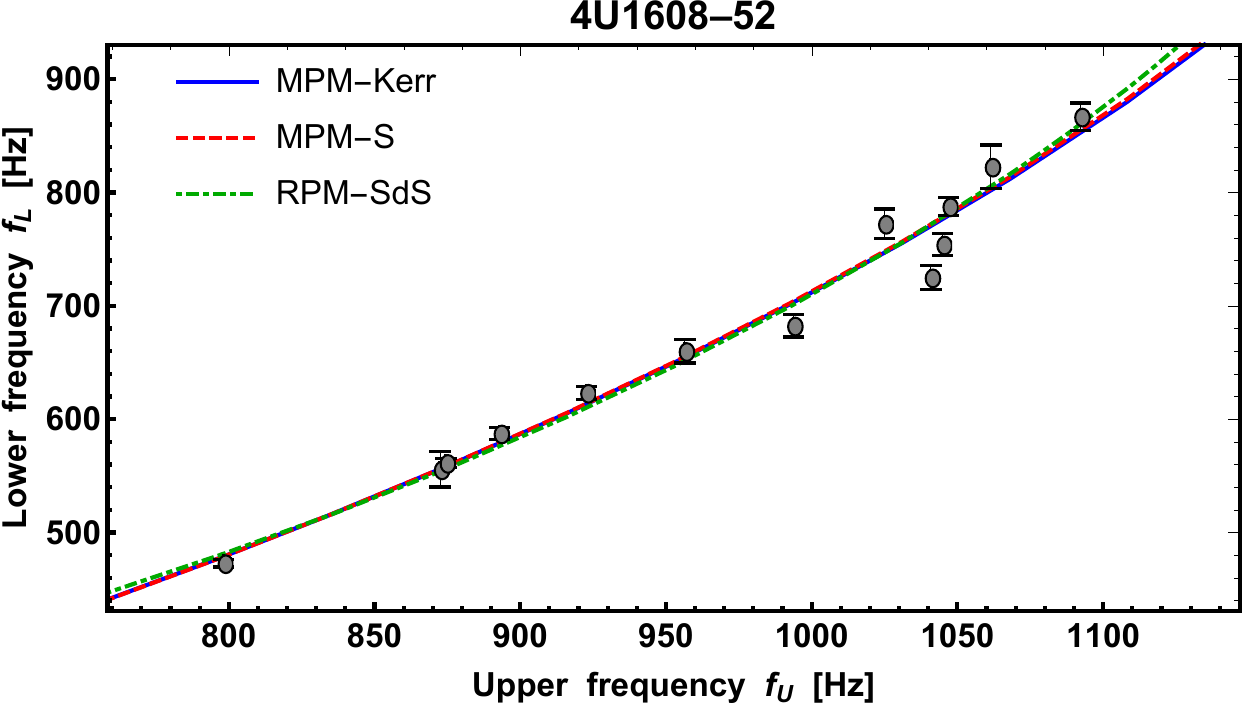}
}

\vspace{0.15cm}
{
\includegraphics[width=0.49\hsize,clip]{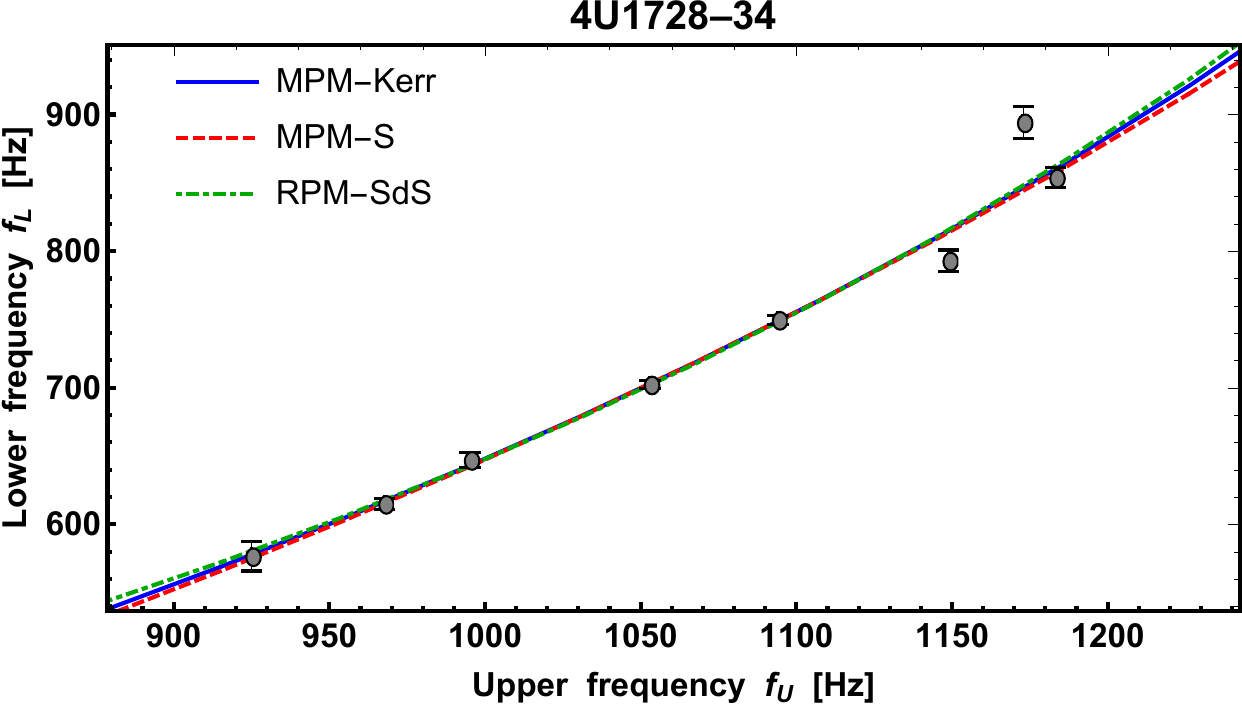}\hfill
\includegraphics[width=0.49\hsize,clip]{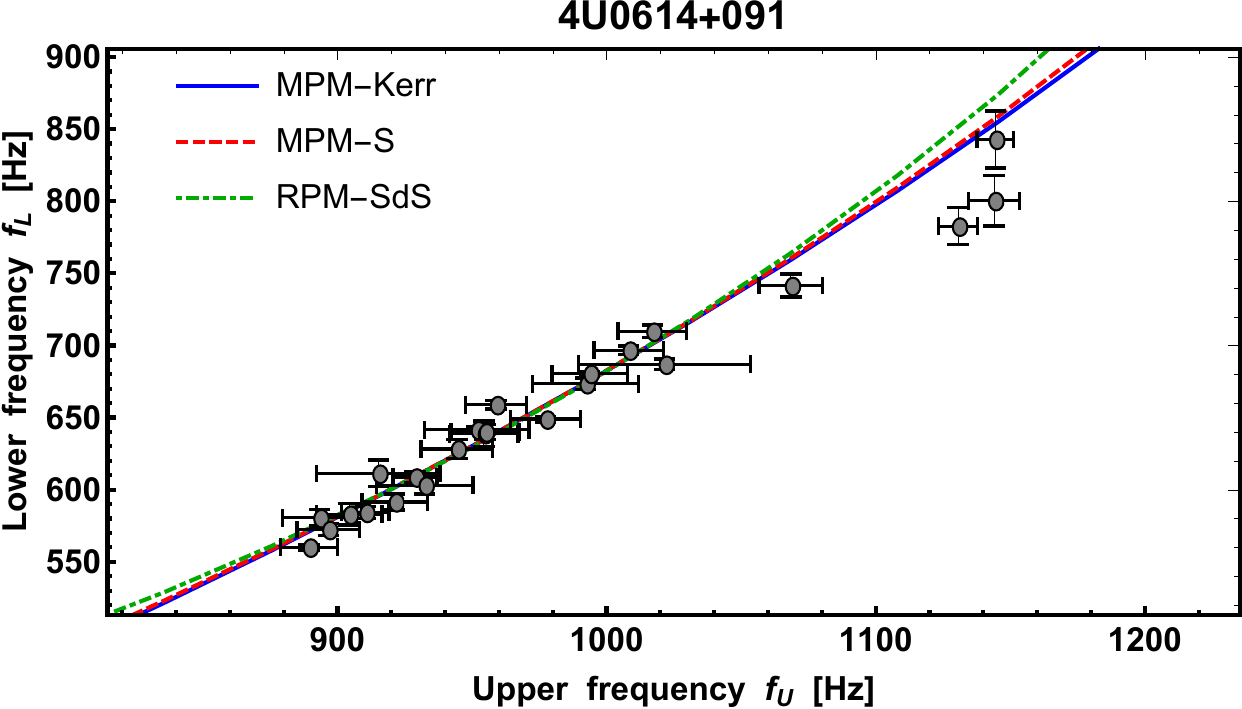}
}\\
\vspace{-0.3cm}
\caption{Lower-upper frequency pairs $(f_L,f_U)$ of the eight NS X-ray binaries and the best-fitting curves for MPM-S (dashed red), MPM-K (solid blue), and RPM-SdS (dot-dashed green).}
\label{fig:freq}
\end{figure}

The fits of the lower-upper frequency pairs $(f_L,f_U)$ of the eight NS X-ray binaries considered in this work are portrayed in Fig.~\ref{fig:freq}. 
The best-fit parameters used to construct the curves shown in Fig.~\ref{fig:freq}
are summarized in Table~\ref{tab:results}, together with the results of the statistical analysis and the reconstruction of the disk structure. 

\begin{table}[t]
\centering
\tiny
\caption{Comparisons of the MCMC best-fit results.}
\setlength{\tabcolsep}{.7em}
\renewcommand{\arraystretch}{1.3}
\begin{tabular}{llrrlrrrrrrr}
\hline\hline
Source                                  &
Model                                   &  
\multicolumn{4}{c}{MCMC best-fit model parameters} &
\multicolumn{2}{c}{Statistical criteria}&
\multicolumn{4}{c}{Characteristic disk radii} \\
&                                       &
$M$                                     & 
$j$                                     &
$n$                                     &
$\mathcal C_n/10^{-4}$                  &
$-\ln \bar L$                           &
$\Delta$                                &
$r_{\rm ISCO}$                          & 
$r_{\rm in}$                            & 
$r_{\rm out}$                           & 
$r_{\rm disk}$                          \\
& & (M$_\odot$) & & & (km$^{1-n}$) & & 
& (km) & (km) & (km) & (km) \\
\hline\hline
                                        &
RPM--S                                  &
$2.224^{+0.029}_{-0.029}$               & 
--                                      &
--                                      &
--                                      &
$125.84$ & $62$                         &
$19.62$ & $30.79$ & $52.16$ & $-$ \\
                                        &
RPM--SdS                                &
$1.846^{+0.045}_{-0.045}$               & 
--                                      &
$3^\star$                               &
$0.13^{+0.01}_{-0.01}$                  &
$70.07$ & $7$                           &
$16.32$ & $28.84$ & $48.29$ & $-$ \\
Cir X1                                  &
RPM--K                                  &
$5.118^{+0.148}_{-0.236}$               &
$>0.97$                                 &
--                                      & 
--                                      &
$77.39$ & $14$                          &
$6.81$ & $60.11$ & $154.08$ & $-$ \\
                                        &
MPM--S                                  &
$1.283^{+0.056}_{-0.058}$  &
--                                      &
$2$                                     &
$18.13^{+0.93}_{-1.19}$                 &
$65.96$ & $3$                           &
$11.92$ & $21.74$ & $29.60$ & $30.44$ \\
                                        &
MPM--K                                  &
$1.944^{+0.767}_{-0.396}$               &
$0.44^{+0.27}_{-0.46}$                  &
$1$                                     &
$-130.89^{+20.32}_{-25.21}$             &
$61.84$ & $0$                           &
$12.85$ & $30.03$ & $52.37$ & unc \\
\hline
                                        &
RPM--S                                  &
$2.161^{+0.010}_{-0.010}$               & 
--                                      &
--                                      &
--                                      &
$200.33$ & $93$                         &
$19.06$ & $21.33$ & $31.70$ & $-$ \\
                                        &
RPM--SdS                                &
$2.397^{+0.019}_{-0.019}$               & 
--                                      &
$3^\star$                               &
$-0.65^{+0.05}_{-0.05}$                 &
$106.08$ & $0$                          &
$20.73$ & $22.15$ & $33.35$ & $-$ \\
GX 5--1                                 &
RPM--K                                  &
$1.286^{+0.027}_{-0.014}$               &
$<-0.94$                                & 
--                                      &
--                                      &
$141.66$ & $36$                         &
$11.53$ & $19.75$ & $49.51$ & $-$ \\ 
                                        &
MPM--S                                  & 
$2.427^{+0.030}_{-0.031}$               &
--                                      &
$3$                                     &
$-0.09^{+0.01}_{-0.01}$                 &
$105.73$ & $0$                          &
$20.91$ & $22.32$ & $34.20$ & $72.68$ \\
                                        &
MPM--K                                  & 
$1.710^{+0.228}_{-0.159}$               &
$-0.39^{+0.18}_{-0.18}$                 &
$1$                                     &
$134.76^{+97.84}_{-69.28}$              &
$104.64$ & $0$                          &
$18.00$ & $19.11$ & $28.18$ & unc \\
\hline
                                        & 
RPM--S                                  &
$2.077^{+0.001}_{-0.001}$               &
--                                      &
--                                      &
--                                      &
$1819.02$ & $1771$                      &
$18.32$ & $18.33$ & $22.94$ & $-$       \\
                                        &
RPM--SdS                                &
$1.733^{+0.011}_{-0.011}$               &
--                                      &
$3^\star$                               &
$2.15^{+0.05}_{-0.05}$                  &
$46.42$ & $0$                           &
$16.01$ & $17.04$ & $21.09$ & $-$       \\
GX 17+2                                 &
RPM--K                                  &
$8.602^{+0.144}_{-0.165}$               &
$>0.99$                                 &
--                                      &
--                                      &
$53.26$ & $7$                           &
$11.31$ & $27.41$ & $41.72$ & $-$ \\
                                        &
MPM--S                                  &
$1.691^{+0.015}_{-0.017}$               &
--                                      &
$3$                                     &
$0.31^{+0.01}_{-0.01}$                  &
$46.56$ & $0$                           &
$15.78$ & $16.79$ & $20.56$ & $41.89$ \\
                                        &
MPM--K                                  &
$2.074^{+0.657}_{-0.234}$               &
$0.28^{+0.14}_{-0.33}$                  &
$1$                                     &
$-813.56^{+\ \,33.52}_{-191.97}$        &
$46.76$ & $2$                           &
$16.31$ & $19.85$ & $25.81$ & $26.25$ \\
\hline
                                        &
RPM--S                                  &
$2.102^{+0.003}_{-0.003}$               & 
--                                      &
--                                      &
--                                      &    
$130.86$ & $5$                          & 
$18.54$ & $21.52$ & $29.07$ & $-$       \\
                                        &
RPM--SdS                                &
$2.149^{+0.015}_{-0.015}$               & 
--                                      &
$3^\star$                               &
$-0.14^{+0.05}_{-0.04}$                 &
$126.06$ & $1$                          &
$18.88$ & $21.71$ & $29.34$ & $-$       \\
GX 340+0                                &
RPM--K                                  &
$1.555^{+0.159}_{-0.123}$               & 
$-0.52^{+0.18}_{-0.17}$                 &
--                                      &
--                                      &
$124.79$ & $0$                          &
$11.84$ & $20.34$ & $45.12$ & $-$ \\
                                        &
MPM--S                                  &
$2.274^{+0.066}_{-0.049}$               &
--                                      &
$2$                                     &
$-4.87^{+1.36}_{-1.84}$                 & 
$124.78$ & $0$                          &
$19.67$ & $22.51$ & $31.09$ & $86.09$ \\
                                        &
MPM--K                                  &
$1.681^{+0.221}_{-0.147}$               &
$-0.34^{+0.18}_{-0.17}$                 &
$1$                                     &
$-34.63^{+55.44}_{-53.56}$              & 
$123.78$ & $0$                          &
$17.50$ & $19.88$ & $27.03$ & unc \\
\hline
                                        &
RPM--S                                  &
$1.965^{+0.001}_{-0.001}$               & 
--                                      &
--                                      &
--                                      &
$3887.17$ & $3754$                      &
$17.33$ & $17.72$ & $20.98$ & $-$      \\
                                        &
RPM--SdS                                &    $1.690^{+0.003}_{-0.003}$               & 
--                                      &
$3^\star$                               &
$2.18^{+0.02}_{-0.03}$                  &
$158.61$ & $27$                         &
$15.60$ & $16.66$ & $19.58$ & $-$ \\ 
Sco X1                                  &
RPM--K                                  &    $6.352^{+0.062}_{-0.068}$               &
$0.93^{+0.01}_{-0.01}$                  &
--                                      &
--                                      &
$131.68$ & $0$                          &
$13.49$ & $25.35$ & $34.90$ & $-$ \\ 
                                        &
MPM--S                                  &
$1.372^{+0.006}_{-0.007}$               &
--                                      &
$2$                                     &
$29.71^{+0.38}_{-0.37}$                 &
$136.72$ & $5$                          &
$13.49$ & $14.42$ & $16.45$ & $23.13$ \\
                                        &
MPM--K                                  &
$1.521^{+0.119}_{-0.103}$               &
$0.12^{+0.07}_{-0.08}$                  &
$2$                                     &
$30.05^{+0.58}_{-0.69}$                 &
$135.94$ & $6$                          &
$13.89$ & $14.93$ & $17.06$ & $23.80$ \\
\hline
                                        & 
RPM--S                                  &
$1.960^{+0.004}_{-0.004}$               & 
--                                      &
--                                      &
--                                      &
$235.83$ & $174$                        &
$17.29$ & $17.65$ & $21.75$ & $-$ \\
                                        & 
RPM--SdS                                &
$1.728^{+0.014}_{-0.014}$               &
--                                      &
$3^\star$                               &
$1.76^{+0.09}_{-0.09}$                  &
$66.14$ & $5$                           &
$15.82$ & $16.77$ & $20.50$ & $-$ \\
4U1608--52                              &
RPM--K                                  &
$5.938^{+0.303}_{-0.259}$               &
$0.91^{+0.02}_{-0.02}$                  &
--                                      &
--                                      &
$60.96$ & $0$                           &
$13.52$ & $24.46$ & $37.84$ & $-$ \\
                                        &
MPM--S                                  & 
$1.429^{+0.027}_{-0.037}$               &
--                                      &
$2$                                     &
$26.08^{+1.89}_{-1.44}$                 &
$62.32$ & $1$                           &
$13.87$ & $14.75$ & $17.42$ & $25.43$ \\
                                        &
MPM--K                                  &       $1.754^{+0.392}_{-0.220}$               &
$0.29^{+0.11}_{-0.21}$                  &
$2$                                     &
$26.52^{+2.23}_{-2.84}$                 &
$62.12$ & $2$                           &
$13.90$ & $15.72$ & $18.67$ & $26.89$ \\
\hline
                                        & 
RPM--S                                  &
$1.734^{+0.003}_{-0.003}$               & 
--                                      & 
--                                      & 
--                                      &
$212.61$ & $177$                        &
$15.30$ & $16.06$ & $18.93$ & $-$ \\
                                        & 
PRM--SdS                                &
$1.445^{+0.016}_{-0.016}$               & 
--                                      & 
--                                      & 
$3.07^{+0.16}_{-0.16}$                  &
$35.15$ & $0$                           &
$13.37$ & $14.96$ & $17.43$ & $-$ \\
4U1728--34                              &
RPM--K                                  &
$6.566^{+0.300}_{-0.323}$               &
$0.98^{+0.01}_{-0.01}$                  &
--                                      &
--                                      &
$36.07$ & $1$                           &
$10.49$ & $24.36$ & $33.69$ & $-$ \\
                                        & 
MPD-S                                   &
$1.115^{+0.032}_{-0.030}$               & 
--                                      &
$2$                                     &
$39.42^{+2.27}_{-2.19}$                 &
$35.02$ & $0$                           &
$11.11$ & $12.42$ & $14.02$ & $17.98$ \\
                                        &
MPM--K                                  &
$1.646^{+0.274}_{-0.229}$               & 
$0.08^{+0.15}_{-0.23}$                  &
$1$                                     &
$-794.03^{+\ \,69.30}_{-113.85}$        &
$34.73$ & $1$                           &
$14.78$ & $17.42$ & $21.08$ & $22.31$ \\
\hline
                                        & 
RPM--S                                  &
$1.904^{+0.001}_{-0.001}$               & 
--                                      &
--                                      & 
--                                      &
$842.97$ & $686$                        &
$16.80$ & $16.95$ & $20.05$ & $-$ \\
                                        & 
RPM--SdS                                &
$1.545^{+0.011}_{-0.011}$               & 
--                                      &
$3^\star$                               &
$2.84^{+0.08}_{-0.08}$                  &
$188.70$ & $33$                         &
$14.34$  & $15.60$ & $18.30$ & $-$ \\
4U0614+091                              &
RPM--K                                  &
$7.708^{+0.043}_{-0.139}$               &
$>0.96$                                 &
--                                      &
--                                      &
$155.32$ & $0$                          &
$9.16$ & $26.42$ & $35.90$ & $-$ \\
                                        &
MPM--S                                  &
$1.154^{+0.019}_{-0.020}$               &
--                                      &
$2$                                     &
$40.07^{+1.22}_{-1.15}$                 &
$170.17$ & $15$                         &
$11.63$ & $12.79$ & $14.45$ & $18.05$ \\
                                        &
MPM--K                                  &
$1.633^{+0.684}_{-0.198}$               &
$0.38^{+0.23}_{-0.14}$                  &
$2$                                     &
$43.25^{+4.31}_{-3.05}$                 &
$168.40$ & $15$                         &
$12.54$ & $14.34$ & $16.29$ & $19.40$   \\ 
\hline\hline
\end{tabular}
\tablefoot{Columns list: the source, the model, the best-fit parameters $M$, $j$, and $\mathcal C_n$ with $1\sigma$ errors and $n$ (fixed), the LLH maximum as $-\ln \bar L$, the Bayes evidence differences $\Delta$, the values of ISCO, inner, outer and, when possible, disk radii. The results of RPM-S and RPM-SdS, taken from \citet{boshkayev23b}, and RPM-K, taken from \citet{2024MNRAS.531.3876B}, are obtained with the same pipeline used here for MPM-S and MPM-K. For RPM-SdS we consider an effective $n=3$ (marked with $\star$) to ensure the correct physical dimensions of the cosmological constant.
The values of the amplitudes $\mathcal C_n$ here reported differ by a factor of $-2$ with respect to the ones in \citet{Bianchini:2025zvp}, due to a different rescaling of the relativistic equations. However, the statistical and physical implications remain unchanged.}
\label{tab:results} 
\end{table}

\end{appendix}

\end{document}